\documentclass[]{relaxed_system_lab}

\usepackage[utf8]{inputenc}
\usepackage[T1]{fontenc}
\usepackage{geometry}
\usepackage{amsmath,amssymb}  %
\usepackage{charter}
\usepackage{varwidth}
\usepackage{wrapfig}
\usepackage{caption}

\usepackage[toc,page,header]{appendix}

\usepackage{minitoc}
\usepackage{cleveref} 
\usepackage{subcaption}
\usepackage{booktabs}
\usepackage{graphicx}
\usepackage{pgfplots}
\usepackage{pgfplotstable}
\usepackage{xcolor}
\usepackage{CJKutf8}
\usetikzlibrary{patterns}
\usepackage{float}
\usepackage{caption}
\usepackage{bm}

\usepackage{soul} %
\usepackage{algorithm}      %
\usepackage{algpseudocode}  %
\usepackage{parskip}

\usepackage{hyperref}
\usepackage{multirow}
\usepackage{url}
\usepackage{caption}
\usepackage{amsmath}
\usepackage{amssymb}
\usepackage{graphicx}
\usepackage{xspace}
\usepackage{xcolor}
\usepackage{tcolorbox}
\usepackage{wrapfig}

\definecolor{lightgreen}{RGB}{144,238,144}
\definecolor{darkcyan}{HTML}{008B8B}
\newcommand{\bluecomment}[1]{{\color{darkcyan}/* #1 */}}
\newcommand{\sys}{\textsc{AReaL-Hex}\xspace}
\newcommand{\sysh}{\textsc{AReaL}\xspace}

\definecolor{cvprblue}{rgb}{0.21,0.49,0.74}
\definecolor{fallbackgreen}{rgb}{130, 180, 102}
\definecolor{stopred}{rgb}{251, 225, 224}

\ifdefined\final
\usepackage[disable]{todonotes}
\else
\usepackage[textsize=tiny]{todonotes}
\fi

\usepackage{natbib}
\usepackage{latexsym}

\usepackage{url}
\usepackage{amssymb}
\usepackage[utf8]{inputenc}
\usepackage{microtype}
\usepackage{booktabs}
\usepackage{pifont} 
\usepackage{multirow}
\usepackage{makecell}
\usepackage{paralist}
\usepackage{xspace}
\usepackage{color}
\usepackage{xcolor}
\usepackage{colortbl}
\usepackage{adjustbox}
\usepackage{hyperref} 
\usepackage[edges]{forest}
\usepackage{tikz} 
\usepackage{caption}
\usepackage{amsfonts}

\hypersetup{
    colorlinks,
    linkcolor={blue!80!black},
    citecolor={blue!80!black},
}
\tikzset{
    root/.style =             {align=center, text width=1cm, rounded corners=3pt, line width=0.3mm, fill=gray!10, draw=gray!80, font=\small},
    demographic/.style =         {align=center, text width=1.8cm, rounded corners=3pt, line width=0.3mm, fill=blue!10, draw=blue!80, font=\footnotesize},
    demographic_work/.style =    {align=center, text width=10cm, rounded corners=3pt, line width=0.3mm, fill=blue!10, draw=blue!0, font=\footnotesize},
    character/.style =         {align=center, text width=1.8cm, rounded corners=3pt, line width=0.3mm, fill=red!10, draw=red!80, font=\footnotesize},
    character_work/.style =    {align=center, text width=10cm, rounded corners=3pt, line width=0.3mm, fill=red!10, draw=red!0, font=\footnotesize},
    personalization/.style =           {align=center, text width=1.8cm, rounded corners=3pt, line width=0.3mm, fill=cyan!10, draw=cyan!80, font=\footnotesize},
    personalization_work/.style =      {align=center, text width=10cm, rounded corners=3pt, line width=0.3mm, fill=cyan!10, draw=cyan!0, font=\footnotesize},
    risk/.style =         {align=center, text width=1.8cm, rounded corners=3pt, line width=0.3mm, fill=orange!10, draw=orange!80, font=\footnotesize},
    risk_work/.style =    {align=center, text width=10cm, rounded corners=3pt, line width=0.3mm, fill=orange!10, draw=orange!0, font=\footnotesize},
}

\usepackage{CJK}

\newtcolorbox{promptbox}[1][]{
  enhanced,
  breakable,
  colback=promptboxlightgray,
  colframe=promptboxblue!30,
  arc=8pt,
  boxrule=0.5pt,
  left=12pt,
  right=12pt,
  top=8pt,
  bottom=8pt,
  fonttitle=\bfseries,
  fontupper=\linespread{1.2}\selectfont,
  title=#1
}

\title{\sys: Accommodating Asynchronous RL Training over Heterogeneous GPUs}

\author{Ran Yan$^1$$^*$, Youhe Jiang$^1$$^*$, Tianyuan Wu$^1$, Jiaxuan Gao$^2$, Zhiyu Mei$^3$, Wei Fu$^2$, Haohui Mai$^1$, Wei Wang$^1$, Yi Wu$^2$, Binhang Yuan$^1$}

\affiliation{$^1$HKUST, $^2$Tsinghua University, $^3$Ant Group}

\contribution{$^*$Equal contribution}

\abstract{
Maximizing the training throughput and cost‑efficiency of reinforcement learning (RL) for large language models (LLMs) is essential to democratize this advanced technique. One promising but challenging approach is to deploy such a computational workflow over heterogeneous GPU clusters.  Unlike conventional large-scale LLM pretraining, RL training generally decomposes into three coupled stages, i.e., rollout generation, reward computation, and policy/value updates, which exhibit markedly different compute intensities, memory footprints, and communication patterns. Recent research shows that fully asynchronous RL training can disaggregate these stages across disjoint hardware pools without sacrificing training stability, creating a great opportunity for real-world heterogeneous deployment. Towards this end, we present \sys, a heterogeneity‑aware asynchronous RL training system that effectively schedules how to execute rollout generation and policy model training over heterogeneous GPU clusters while enforcing data staleness bounds. Concretely, \sys uses a two‑phase scheduler: (\underline{\textbf{i}}) a constrained search with mixed‑integer linear programming to select per‑stage parallelization strategies and workload assignments given a resource budget, and (\underline{\textbf{ii}}) a graph‑partitioning step that allocates heterogeneous GPUs and interconnects to maximize end‑to‑end throughput. Built atop a fully asynchronous RL architecture, \sys maps HBM‑I/O‑bound generation and compute‑bound optimization to more cost-efficient resources and balances their producer–consumer interactions to avoid both idleness and stale rollout trajectories. On the mathematical reasoning task with various model scales (1.5B, 7B, and 14B), compared to homogeneous deployments of state‑of‑the‑art asynchronous RL systems: (\underline{i}) When maintaining the same total budgets, \sys delivers up to 1.50$\times$ higher training throughput; (\underline{ii}) When achieving the same training throughput, \sys results in up to 1.46$\times$ reduction in training cost.
}

\begin{document}

\maketitle

\section{Introduction}
\label{sec:intro}

Reinforcement learning (RL) has emerged as a popular approach for enhancing large language models (LLMs) across various tasks~\cite{zhang2025survey}, including math~\cite{ren2025deepseek}, code generation~\cite{hui2024qwen2}, agentic tool use~\cite{yao2025tau}, etc, which becomes essential to advance the development of AI. Unlike conventional large-scale pretraining by standard stochastic gradient-based optimization, RL training comprises multiple coupled stages, where each stage includes heterogeneous compute intensities, memory footprints, and communication patterns~\cite{sheng2025hybridflow}; furthermore, recent study suggests that asynchronous RL paradigm can effectively disaggregate these different stages over different set of hardwares without compromising the stability in RL training~\cite{fu2025areal,zhong2025streamrl}. Such an intrinsic computational heterogeneity under the disaggregated infrastructure naturally provides a great opportunity to deploy such a workflow over a heterogeneous GPU cluster. In this paper, we explore \textit{how to effectively schedule asynchronous RL workloads on heterogeneous resources to maximize end-to-end RL training throughput and cost-efficiency}.

\begin{figure*}
    \centering
    \includegraphics[width=\linewidth]{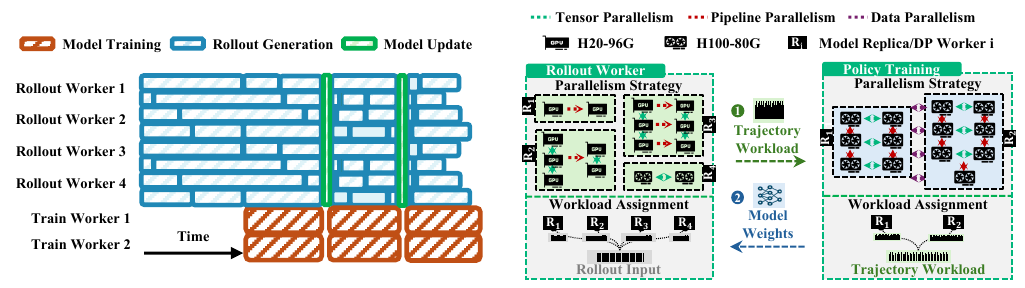}

    \caption{\underline{\textbf{Left:}} Illustration of the asynchronous RL training workflow. The computations for rollout generation and model training are overlapped. When a new version of model weights becomes available, both rollout and training processes are temporarily paused to update the rollout workers’ weights. \underline{\textbf{Right:}} An example of scheduling results. Multiple rollout replicas, potentially configured with distinct parallelization strategies and hardware resources, are created to perform rollout generation tasks. (\textbf{\underline{i}}) Once rollouts are completed, they are transmitted to the training workers. (\textbf{\underline{ii}}) When the updated model weights are ready, training workers broadcast the new weights to rollout workers via NCCL collective communication operations.}

    \label{fig:system-overview}
\end{figure*}

Deploying asynchronous RL training in heterogeneous environments promises substantial gains in resource utilization and reductions in service cost. RL training for LLM is highly resource‑intensive and includes three different stages: (\underline{\textbf{i}}) \textit{rollout generation}, which conducts HBM IO–bound inference on GPUs that produces multiple responses (known as rollout) per prompt; (\underline{\textbf{ii}}) \textit{reward estimation}, which uses strategies that may require large numbers of CPU cores (e.g., sandboxed execution for coding tasks or rule‑based evaluation for math) or additional GPU capacity for LLM‑based value models; and (\underline{\textbf{iii}}) \textit{model training}, which is compute‑bound GPU updates of the policy (and optionally value) LLMs via stochastic gradient‑based optimizations, potentially with a reference model for stability. 
Recently, fully asynchronous RL training paradigm has been shown to be effective, where the three stages can execute over disjoint GPU pools, and the rollout worker GPU keeps the generation continuously while training GPUs update opportunistically, rather than waiting for the longest rollout sequence in a synchronous batch.  
This multi-stage disaggregated paradigm exhibits inherent heterogeneity in terms of compute intensity, memory bandwidth, and communication paradigm in corresponding parallel execution for each stage, making it a natural fit for clusters with different GPU with the heterogeneous GPU-interconnects, which are common in cloud and enterprise settings. We believe \textit{an efficient scheduler} that is able to map the distributed execution in each stage to more cost-efficient devices and balances their asynchronous interactions could fully exploit available heterogeneous hardware, lower both time and monetary cost, and make large‑scale RL more accessible.

However, scheduling asynchronous RL training workflows over heterogeneity is also fundamentally challenging due to the complex training dynamics of RL and the constraints it imposes on the data staleness~\cite{noukhovitcha2025synchronous}. The state-of-the-art asynchronous RL training paradigm operates as a producer–consumer paradigm in which the rollout worker GPUs use the policy LLM to continuously generate rollouts while the training worker GPUs concurrently update the policy LLM. In such an asynchronous situation, an effective heterogeneity-aware scheduler must regulate this interaction under the explicit data staleness constraints, i.e., the training worker GPUs require a steady supply of rollouts, yet such rollouts must not be generated by policy LLM too stale that the rollout trajectory becomes obsolete.
The difficulty is further amplified by the heterogeneity that enlarges the search space: GPU devices differ in compute throughput, HBM memory capacity and bandwidth, and interconnect topology, turning computation task placement into a complicated search problem in an exponential search space. The scheduled plan must simultaneously maintain high GPU device utilization and limit rollout data staleness, which demands some careful design  of the scheduler.

Despite growing interest in heterogeneity‑aware scheduling for LLM serving, existing efforts address only isolated slices of the problem, i.e., either inference or training, thus falling short of the end‑to‑end requirements of asynchronous RL training. For the LLM training workloads, heterogeneous LLM systems such as HexiScale~\citep{yan2024hexiscale} and Metis~\citep{um2024metis} optimize synchronous LLM training across heterogeneous GPUs; on the inference side, serving frameworks including Helix~\citep{mei2025helix}, HexGen~\citep{jiang2024hexgen}, and HexGen‑2~\cite{jiang2025hexgen2} orchestrate largely general inference workloads on heterogeneous clusters. Simply composing these techniques is inadequate: an asynchronous RL training pipeline is inherently multi‑stage and requires explicit control of rollout data staleness, ensuring that rollouts consumed by the training worker GPUs are bounded by the configured data staleness constraint. Thus, an effective scheduler must therefore jointly consider cross‑stage heterogeneity and inter‑stage coupling, which makes the scheduling space of asynchronous RL training fundamentally departs from synchronous LLM training or general LLM inference.

In this paper, we introduce \sys, a novel asynchronous RL training system designed to maximize end-to-end training throughput and cost-efficiency for RL training workloads on heterogeneous GPU clusters. 
The core of \sys is a heterogeneity-aware scheduling algorithm that determines effective allocation of generation and training computation and corresponding communication under different parallelism across heterogeneous GPUs and network interconnects. Concretely, we summarize our contributions as follows: 

\begin{itemize}

    \item \textbf{\underline{Contribution 1.}}
    We systematically characterize multi-stage asynchronous RL workloads and demonstrate that their intrinsic computational heterogeneity, i.e., HBM IO-bound rollout generation versus compute-bound policy model training, naturally aligns with heterogeneous GPU clusters. Through extensive profiling, we show that matching each stage's resource demands to appropriate GPU types can substantially improve overall system efficiency compared to homogeneous deployments.
\vspace{-0.25em}
    \item
    \textbf{\underline{Contribution 2.}}
    We formulate the problem of asynchronous RL training scheduling over heterogeneity, which encompasses resource allocation, parallel strategies, and workload assignments, as a constrained optimization problem with explicit data staleness bounds. To efficiently solve the scheduling problem, we propose a two-phase algorithm: (\underline{\textbf{i}}) the first phase employs constrained search and mixed-integer linear programming (MILP) to determine optimal parallel strategies and workload assignments for generation and training under a given resource allocation; and (\underline{\textbf{ii}}) the second phase uses graph partitioning to optimize resource allocation across heterogeneous devices, maximizing end-to-end training performance.
\vspace{-0.75em}
    \item 
    \textbf{\underline{Contribution 3.}}
    We implement \sys and conduct extensive experiments for mathematical reasoning RL tasks and model scales (1.5B, 8B, and 14B). We want to highlight that compared to homogeneous deployments of state-of-the-art asynchronous RL systems, \sys achieves 1.31$\times\sim$1.50$\times$ improvement in training throughput given the same training budget, demonstrating the substantial benefits of heterogeneity-aware scheduling for asynchronous RL training. On the other hand, when achieving the same training throughput, \sys results in up to 1.46$\times$ reduction in training cost. 

\end{itemize}

\section{Preliminaries and Related Work}
\label{sec:preliminaries}

\subsection{RL Training System for LLMs.}
\vspace{-0.5em}

RL has been applied to enhance the reasoning capabilities of LLMs~\cite{openai_learning_to_reason_llms_2024,openai_o3_o4mini_2025}. 
Existing RL training demonstrates strong performance on reasoning tasks, achieving significant improvements in various tasks, including mathematical problem solving, code generation, and multi-hop question answering~\cite{chen2021evaluating,hendrycks2021measuring,liu2023agentbench,wang2025reinforcement,gao2025beyond}. 
RL training includes both the HBM IO-bound rollout generation phase and the compute-bound training phase~\cite{zhong2025optimizing,zhong2025streamrl,fu2025areal}, where both phases require careful orchestration of different parallelism strategies (e.g., data-, pipeline-, and tensor model- parallelisms), GPU resource allocation, and sequence-level scheduling (e.g., packing, KV cache management)~\cite{li2023alpaserve,huang2019gpipe,shoeybi2019megatron,miao2022galvatron,jiang2023osdp}. RL training for LLMs can be categorized into synchronous and asynchronous paradigms.
In \textit{synchronous} RL training, different phases are iteratively executed: the rollout generation phase uses the latest model parameters to produce reasoning traces for each input question, and the training phase updates the model parameters based on these rollouts~\cite{schulman2017proximal,sheng2025hybridflow}.
On the other hand, in \textit{asynchronous} RL training, these phases can run in parallel: the generation phase continuously produces rollouts using potentially stale model parameters, and the training phase updates the parameters based on such rollouts~\cite{mnih2016asynchronous,espeholt2018impala,espeholt2019seed,mei2023srl,fu2025areal}.

Due to the distinct computational characteristics of RL training with a multi-stage pipeline, recent efforts attempt to develop efficient RL training systems~\cite{mei2024real,zhong2025optimizing,wu2025g}. Among them,
Verl~\cite{sheng2025hybridflow} accelerates RL training with a 3D-HybridEngine for zero-redundancy resharding between actor training and generation, and an auto-mapping algorithm for GPU allocation and placement.
StreamRL~\cite{zhong2025streamrl} addresses the pipeline bubbles and skewness-induced inefficiencies present in existing disaggregated RL frameworks, enabling flexible resource allocation. AsyncFlow~\cite{han2025asyncflow} allows one-step parameter asynchronization between rollout generation and policy model update, minimizing hardware idling across iterations while maintaining algorithmic integrity.
AReaL~\cite{fu2025areal} decouples streaming generation from training through a fully asynchronous architecture and incorporates several algorithmic innovations, including staleness-aware training and a decoupled RL objective, which enables efficient and stable RL training for LLM reasoning.
Different from existing works, \sys focuses on effectively scheduling multi-stage asynchronous RL training workflows across heterogeneous GPU clusters in order to maximize resource utilization and minimizing training costs.


\subsection{LLM Service over Heterogeneity.} 

Earlier cluster management frameworks laid some groundwork for heterogeneous resource scheduling in distributed computation~\cite{hindman2011mesos,lee2011heterogeneity}. For instance, Mesos~\cite{hindman2011mesos} pioneered a two-level scheduling architecture that enables fine-grained sharing of cluster resources among diverse frameworks, while Paragon~\cite{delimitrou2013qos} provided a QoS-aware scheduler that classifies incoming workloads to places them on the most appropriate servers based on hardware differences. Collectively, these efforts explored adapting scheduling policies to heterogeneous infrastructures.

Recent research has investigated methods for leveraging heterogeneous computational resources to enhance the cost-efficiency of LLM services~\cite{um2024metis,mo2025hetis,mao2025skyserve,jiang2025demystifying,jiang2025thunderserve,tong2025parallax,peng2025hexgen}. For instance, several works~\cite{yuan2022decentralized,miao2023sdpipe} explore training LLMs on heterogeneous GPUs to enhance cost-effectiveness and resource utilization. Among them, HexiScale~\cite{yan2024hexiscale} supports hybrid and asymmetric parallelism implementations to leverage heterogeneous GPU capabilities, and optimizes these asymmetric computations with a hierarchical graph partitioning algorithm. In the context of LLM inference and serving, Helix~\cite{mei2025helix} formulates heterogeneous GPU allocation and network connectivity as a max-flow problem, applying mixed-integer linear programming to identify optimal strategies for LLM inference serving. HexGen~\cite{jiang2024hexgen} introduces asymmetric partitioning and advanced scheduling mechanisms for generative inference in decentralized, heterogeneous settings. HexGen-2~\cite{jiang2025hexgen2} further extends the heterogeneity-aware scheduling on top of the disaggregated inference paradigm. 

\section{Observation and Opportunity}
\label{sec:case}

Asynchronous RL training exhibits intrinsic computational heterogeneity, where the HBM I/O-bound rollout generation and compute-bound model training can be disaggregated over different sets of GPUs. This disaggregated computational heterogeneity motivates the design of \sys: \textit{Heterogeneous clusters can naturally align their different hardware capabilities with the heterogeneous computational demands of asynchronous RL training, providing opportunities to reduce the real-world training costs.} We explore this motivation and present our observations through an illustrative case study.

\textbf{Benchmark setups.} We maintain the same total budget, and benchmark the asynchronous RL training throughput across three settings on the 1.5B, 7B, and 14B models: (\textbf{\underline{i}}) \textit{Homogeneous setting 1}: 32 H800 GPUs. (\textbf{\underline{ii}}) \textit{Homogeneous setting 2}: 88 H20 GPUs. (\textbf{\underline{iii}}) \textit{Heterogeneous setting}: 24 H800 GPUs and 24 H20 GPUs. We benchmark both the rollout generation latency and model training latency, which comprise the major execution cost of RL training. We use the reported H800 and H20 GPU rental price by MegaScale-Infer~\cite{zhu2025megascale}---For the H800 GPU, the rental price is \$5.28 per hour; and for the H20 GPU, the rental price is \$1.85 per hour.

\begin{figure}[t!]
    \centering
    \includegraphics[width=0.8\linewidth]{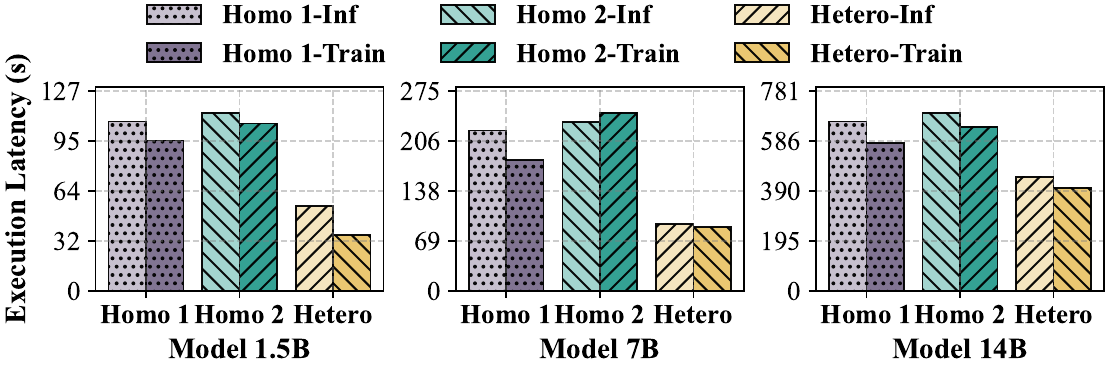}
    
    \caption{Rollout inference (short as \textsc{Inf}) and model training (short as \textsc{Train}) execution latency comparison between homogeneous setting 1, 2 and heterogeneous setting across different model scales.}
    \label{fig:profile}
    
\end{figure}

\textit{Observation 1: H800 GPUs with high computational power are not efficient for inference.} H800 GPUs have limited HBM memory bandwidth; to avoid being bound by rollout generation, we must allocate a large fraction of H800 GPUs to rollout generation while allocating only a small fraction to model training. Under asynchronous execution, although we can balance rollout generation and model training throughput, the overall performance remains suboptimal due to inefficient rollout generation performance.

\textit{Observation 2: H20 GPUs are insufficient for computationally intensive model training.} Homogeneous setting 2 falls short in compute-bound model training, as the H20 GPUs have limited computational capabilities in terms of peak FLOPS. Since H20 GPUs provide only limited computational power, we must allocate a large fraction of H20 GPUs to model training. Additionally, training with a large number of H20 GPUs introduces scalability issues; empirically, we find that training with 5$\times$ H20 GPUs cannot sufficiently achieve the same training throughput as H800 GPUs, even if the theoretical computation power of 5 H20 GPUs is equal to 1 H800 GPU. As a result, the overall performance of homogeneous setting 2 remains suboptimal due to inefficient model training.    

\textit{Observation 3: Heterogeneous setting naturally leverages the advantages of both GPU types.} The heterogeneous setting, which hybridizes compute-efficient H800 GPUs and memory bandwidth-efficient H20 GPUs, naturally leverages the advantages of both GPU types, thereby demonstrating superior system performance. We disaggregate the rollout generation and model training on different hardware. On the one hand, we assign the HBM memory-bound rollout generation phase to H20 GPUs, fully utilizing the abundant HBM memory bandwidth. On the other hand, we assign the compute-bound model training phase to H800 GPUs, fully leveraging the advanced computational power. Under this arrangement, both phases achieved optimal performance, thereby boosting the system performance. Compared with the homogeneous settings 1 and 2, the heterogeneous setting achieved up to 2.67$\times$, and at least 1.49$\times$ reduction in the end-to-end RL training execution time.  

\begin{table}[h]
\centering

\caption{Rollout and training per-token cost comparison of different GPU types on different model scales.}

\label{tab:per-token-cost}
\resizebox{0.6\columnwidth}{!}{%
\begin{tabular}{l | l | c | c}
\hline
\textbf{Model} & \textbf{GPU} & \textbf{\$ Per Inference Token} & \textbf{\$ Per Training Token} \\
\hline
\hline
\multirow{2}{*}{\textbf{1.5B }} & H800 & \$$5.58\times10^{-3}$ & \$$1.66\times10^{-3}$ \\
\cline{2-4}
 & H20 & \$$2.06\times10^{-3}$ & \$$5.16\times10^{-3}$ \\
\hline
\multirow{2}{*}{\textbf{7B }} & H800 & \$$1.14\times10^{-2}$ & \$$3.05\times10^{-3}$ \\
\cline{2-4}
 & H20 & \$$4.15\times10^{-3}$ & \$$9.51\times10^{-3}$ \\
\hline
\multirow{2}{*}{\textbf{14B }} & H800 & \$$3.43\times10^{-2}$ & \$$0.99\times10^{-2}$ \\
\cline{2-4}
 & H20 & \$$1.27\times10^{-2}$ & \$$3.10\times10^{-2}$ \\
\hline
\end{tabular}%
}
\end{table}

\textbf{Cost-efficiency analysis of different GPU types.} Asynchronous RL training workloads across heterogeneous disaggregated GPUs naturally align with the distinct cost-efficiency of different GPU types. In \autoref{tab:per-token-cost}, we present the per-token cost for H800 and H20 GPUs under both training and inference computation. For rollout generation, H20 demonstrates 2.72$\times$ higher cost efficiency; for training, H800 demonstrates 3.12$\times$ higher cost efficiency. These results confirm that the performance gains from hybridizing different GPUs originate from utilizing more cost-efficient devices for the two phases of RL training.

\section{Scheduling with Heterogeneity}

We first introduce the scheduling problem and the overview of our two-phase solution in\S\ref{sec:problem} and then propose introduce the search phase in \S\ref{sec:first-phase} and repartition phase in \S\ref{sec:second-phase}. 

\subsection{Problem Formulation}
\label{sec:problem}

\noindent \textbf{Formulation.} Given a set of heterogeneous GPUs noted by $\mathcal{D}$, staleness constraint parameterized by $\eta$, and an expected generated rollout length distribution $P$, the goal of the scheduler is to identify the optimal asynchronous RL execution plan that minimizes the $\delta(\eta)$-step averaged RL training time, where $\delta(\eta)$ represents a window size used to calculate the average training time under asynchronous rollout generation, depending on the staleness $\eta$. Concretely, to optimize asynchronous RL training, the scheduling algorithm should determine two essential components:

\begin{itemize}
     \item \textit{Resource allocation}, which determines the optimal allocation of the GPU set for policy (and optionally value) model training, noted by $\mathcal{D}_T$, where $\mathcal{D}_T \subset \mathcal{D}$ and the GPU set for rollout generation by generative inference noted by $\mathcal{D}_I$, where $\mathcal{D}_I \subset \mathcal{D}$. Note that $\mathcal{D}_T \cup \mathcal{D}_I = \mathcal{D}, \mathcal{D}_T \cap \mathcal{D}_I = \emptyset$, and the ideal balances between $\mathcal{D}_T$ and $\mathcal{D}_I$ should maximize overall RL training throughput.
\vspace{-0.25em}
    
    \item \textit{Parallel strategy and workload assignment}, that determines the allocation of the distributed computational workload to each heterogeneous GPU under asynchronous RL training, given a particular resource allocation. To be more specific, this component needs to assign the parallel inference execute plan $\tau$ over the GPU set $\mathcal{D}_I$ and the training plan $\sigma$ over the GPU set $\mathcal{D}_T$.

\end{itemize}%

\noindent We define a solution including these two components as a \textit{scheduled plan} as the output. Concretely, given $\mathcal{D} = \{d_1, \ldots, d_N \}$ as a set of $N$ GPUs, where we note the device memory limit $M_d$ for GPU $d$, we define the scheduling problem to seek optimal parallel execution plans $\sigma^*$ for model training and $\tau^*$ for rollout generation over the optimal resource allocation noted by $\mathcal{D}^*_T$ and $\mathcal{D}^*_I$ that utilize all $N$ GPUs in $\mathcal{D}$ to minimize $\delta(\eta)$-step average RL training execution time subject to each GPU's memory constraints:%

\begin{equation}
\begin{aligned}
&\sigma^{*}, \tau^*, \mathcal{D}^*_T, \mathcal{D}^*_I = \arg \min_{\sigma,\tau, \mathcal{D}_T, \mathcal{D}_I} \max \{ C_T, C_I \} \\
&\text{where} \enspace C_T = C_{\textsc{Train}}\left( \sigma, \mathcal{D}_T, \delta(\eta) \right), \\
&\phantom{\text{where}} \enspace C_I = C_{\textsc{Rollout}}\left(\tau, \mathcal{D}_I, P, \delta(\eta) \right) + C_{\textsc{Reward}}\left(\delta(\eta) \right) + C_{\textsc{Update}}(\sigma, \mathcal{D}_T, \tau, \mathcal{D}_I, \delta(\eta)),\\
&\text{s.t.} \enspace \mathcal{D}_T \cup \mathcal{D}_I = \mathcal{D}, \mathcal{D}_T \cap \mathcal{D}_I = \emptyset, \textsc{Mem-Cumsum}\left( d \right) \leq M_d \quad \forall d \in \mathcal{D}.
\end{aligned}
\end{equation}%

\noindent where $C_T$ represents the $\delta(\eta)$-step average training execution cost, including $C_{\textsc{Train}}(\sigma, \mathcal{D}_T, \delta(\eta))$ that captures the gradient computation and parameter update for policy (optionally value) model; 
$C_I$ represents the $\delta(\eta)$-step average rollout cost, including $C_{\textsc{Rollout}}\left(\tau, \mathcal{D}_I, P, \delta(\eta) \right)$ for rollout generation; $C_{\textsc{Reward}}\left(\delta(\eta) \right)$ for reward computation (involving reward and reference model computation) latency; and $C_{\textsc{Update}}(\sigma, \mathcal{D}_T, \tau, \mathcal{D}_I, \delta(\eta))$ that represents the cost of synchronizing updated policy model weights to rollout workers. . $\textsc{Mem-Cumsum}(d)$ denotes memory consumption on device $d$ to execute the allocated policy $\sigma$ and $\tau$.
We want to be explicit that our algorithm assumes rollout generation execution cost generally exceeds the model training execution cost (i.e., $C_I > C_T$). Because when model training execution cost exceeds rollout generation execution cost (i.e., $C_T > C_I$), the system gradually approaches its data staleness tolerance threshold, causing rollout workers to stall and wait for slow model training before resuming rollout generation after at least one training batch is consumed---In this scenario, computational resources on rollout workers are significantly wasted, degrading overall system throughput. Conversely, when the rollout generation execution cost exceeds the model training execution cost (i.e., $C_I > C_T$), the system can leverage the asynchronous rollout feature to minimize the stalling time of model training. Therefore, we exclude cases where the model training execution cost exceeds the rollout generation execution cost.

\noindent \textbf{Overview of the scheduling algorithm.} Notice that finding the optimal scheduled plan is NP-hard due to the exponential space of candidate allocations. Furthermore, one can imagine that the process of determining these two essential components (i.e., \textit{resource allocation} $\mathcal{D}_T$, $\mathcal{D}_I$ and \textit{parallel strategy and workload assignment $\sigma$, $\tau$}) introduces a ``chicken and egg'' dilemma---the optimal plan $\sigma$, $\tau$ can only be estimated over the optimal resource allocation $\mathcal{D}_T$, $\mathcal{D}_I$; while the resource allocation $\mathcal{D}_T$, $\mathcal{D}_I$ should be determined by the underlying optimal potential plan $\sigma$, $\tau$. To resolve this issue, we propose a two-phase scheduling approach enlightened by the classic EM algorithm~\cite{moon1996expectation}, summarized as below:

\begin{algorithm}[t]
\caption{Scheduling Algorithm}
\label{alg:scheduling}
\begin{algorithmic}

\Statex {\bfseries Input:}
$(\mathcal{D}_T^0, \mathcal{D}_I^0)$: initial allocation, $\mathcal{D}$: GPU device info, 
\Statex \quad \quad \quad $K$: stable iteration threshold, $\delta(\eta)$: window size

\Statex \bluecomment{$\sigma$: model training plan, $\tau$: rollout generation plan}
\Statex {\bfseries Output:} optimized $\sigma^*$, $\tau^*$, $\mathcal{D}^*_T$, $\mathcal{D}^*_I$

\Statex {\bfseries Intermediate Variables:}
\Statex \quad $C_T$: Model training cost, $C_I$: Rollout generation cost

\Statex \bluecomment{Initialize training and inference resource allocation.}
\Statex $\mathcal{D}_T, \mathcal{D}_I \leftarrow \mathcal{D}^0_T,\mathcal{D}^0_I $

\While{convergence not reached}
    \Statex \bluecomment{Search-Phase: Optimize execution plans (\S\ref{sec:first-phase})}
    \Statex $(\sigma, C_T) \leftarrow \textsc{Constrained\_Search}(\mathcal{D}_T)$
    \Statex $(\tau, C_I ) 
    \leftarrow \textsc{MILP}(\mathcal{D}_I, P, \delta(\eta))$
    
    \Statex \bluecomment{Repartition-Phase: Optimize resource allocation (\S\ref{sec:second-phase})}
    \Statex $(\mathcal{D}_T, \mathcal{D}_I) \leftarrow \textsc{Graph\_Partition}(C_T, C_I, \mathcal{D})$
    
    \Statex \bluecomment{Check convergence condition.}
    \If{$\max\{C_T, C_I\}$ is stable for $K$ iterations}
        \Statex {\bfseries break}
    \EndIf
\EndWhile
\end{algorithmic}
\end{algorithm}

\begin{itemize}

\item \textbf{Search-Phase}:
Given a particular fixed resource allocation, we employ \textit{constrained and MILP-based search} to determine the optimal parallel strategy and workload assignment (\S\ref{sec:first-phase}). This phase optimizes system performance for both model training and rollout generation within the allocated resources, generating execution plans with minimum cost.
\vspace{-0.25em}
\item \textbf{Repartition-Phase}:
Given the estimated execution costs obtained from the Search-Phase, we apply a \textit{cost-guided graph partition algorithm} to identify the optimal resource allocation (\S\ref{sec:second-phase}). This phase balances computational resources between model training and rollout generation, ensuring globally optimized system performance.

\end{itemize}%
These two phases alternate iteratively and terminate when the maximum execution cost (i.e., $\max\{C_T, C_I\}$) remains stable for multiple consecutive iterations (e.g., 20 iterations). We summarize this process in Algorithm~\ref{alg:scheduling}.

\subsection{Search Phase Algorithm}
\label{sec:first-phase}

Given a particular resource allocation $\mathcal{D}_T$ and $\mathcal{D}_I$, the search phase determines the optimal execution plans for model training $\sigma$ and rollout generation $\tau$ that maximize their training or inference throughput. 

\subsubsection{Searching Model Training Execution Plan}

Given the resources set $\mathcal{D}_T$ allocated for the model training, we determine an optimal training execution plan $\sigma$, which involves both parallel strategy and workload assignment noted as $\textsc{Constrained\_Search}(\mathcal{D}_T)$, as follows:




\textbf{Parallel strategy search.} Given the heterogeneous resources allocated to model training, we propose an efficient and effective constrained search algorithm to determine an optimal parallel strategy i.e., a combination of data parallelism (DP), tensor model parallelism (TP), and pipeline parallelism (PP), that minimizes training execution cost. 
Our approach is motivated by the observation that different GPU types are typically connected via low-bandwidth networks in real-world distributed environments~\cite{um2024metis,yan2024hexiscale}. To mitigate extensive cross-GPU-type communication overhead, we impose a simple yet effective practical constraint: Both TP and DP must utilize GPUs of the \textit{same} type. Under this constraint, we systematically enumerate all feasible parallel strategies. 
This constraint substantially reduces the search complexity while maintaining solution quality. Finally, we follow the search algorithm in Metis~\cite{um2024metis} to select the parallel strategy that yields the maximal training throughput across all evaluated configurations. The scheduled results will ensure that across pipeline stages, we allocate transformer layers proportionally to the computational capabilities of heterogeneous devices, where each pipeline stage is configured with distinct data and tensor model parallelism.


\textbf{Workload assignment optimization.} We employ a greedy sequence packing strategy~\cite{fu2025areal} to balance token distribution across data parallel workers and optimize GPU memory utilization: For each training batch, we sequentially assign sequences to the DP worker with the minimum current workload, measured by token count.


\subsubsection{Searching Rollout Generation Execution Plan}

Let $\mathcal{D}_I$ be the set of GPUs allocated for rollout generation, we formulate how to find the optimal rollout generation plan as a mixed integer linear program problem (MILP) noted as $\textsc{MILP}(\mathcal{D}_I, P, \delta(\eta))$ as below.

\textbf{Model replica configurations.} 
Different from model training, multiple rollout inference replicas should work in parallel to generate rollouts independently. Note that such replicas could duplicate the same configuration, i.e., execute the same parallel strategy over the same type of GPUs connected by the network with the same bandwidth. To enumerate this process, we include some new notations:
Suppose there are $T$ different types of GPUs, define $\mathcal{T}_I = [i_1, i_2, ..., i_T ]$ as the count of each type of GPUs, e.g., $i_t$ represents the count of type-$t$ GPUs. To determine the execution plan $\tau$, we characterize rollout replica parallel strategies, replica counts, and workload assignment as follows:
    


Let $\Psi$ denote possible configurations for one rollout inference replica. Each configuration $\psi \in \Psi$ represents the configuration for a rollout inference replica characterized by $(v_{\psi}, s_{\psi}, h_{\psi}, y_{\psi})$: (\underline{\textbf{i}}) Vector $v_{\psi}=[i^{\psi}_1, i^{\psi}_2, ..., i^{\psi}_T ] $, $0 \leq i^{\psi}_t \leq i_t, \forall t \in [1, T]$ indicates GPU count of each type indexed by $t$ used in configuration $\psi$. (\underline{\textbf{ii}}) Vector $s_{\psi} =[\text{tp}^{\psi}_1, \text{tp}^{\psi}_2, ..., \text{tp}^{\psi}_{S^{\psi}}]$ indicates this inference serving replica is equipped with $S^{\psi}$ pipeline stages, where $\text{tp}^{\psi}_s$ is the TP degrees of the $s$-th stage ($s\in [1,S^{\psi}]$). (\underline{\textbf{iii}}) The estimated inference throughput $h_{\psi}$ is estimated based on the parallel strategy following the cost model from HexGen~\cite{jiang2024hexgen}. (\underline{\textbf{iv}}) Free integer variable $y_{\psi}$ suggests the number of replicas following configuration $\psi$.



\textbf{Workload assignment to rollout replicas.} 
During rollout generation, the input prompt length is predefined by the task; in contrast, the output length of each rollout is nondeterministic during RL training. Here, we sample the output length from the skew distribution $P$ that can be profiled during the cold-start of RL training to estimate the output tokens, where we note the output sequence length $\text{len}_{\text{rollout}} \sim P$. The allocation problem should determine the number of rollouts that each rollout inference replica should host in its local inference batch. Formally, let $X_{\delta(\eta)}$ denote the maximum processing capacity (measured in number of rollouts) within $\delta(\eta)$ training steps, and let $B$ denote the total number of rollouts that all of the rollout replicas should process. 
Our algorithm determines an assignment $\mathcal{A}_I = \{x_{\psi} : \psi \in \Psi \}$, where $x_{\psi} \in [0, B]$ represents the number rollouts assigned to configuration $\psi$, subject to the constraints that $\sum_{\psi \in \Psi} x_{\psi} = B$.


\textbf{Optimization objective.} 
We define a makespan variable $\Theta \geq 0$ that represents the overall rollout generation completion time across $\delta(\eta)$ training steps. For each configuration $\psi \in \Psi$ that is duplicated $y_\psi$ times and processes $x_\psi$ rollout sequences from the total rollout count $B$, the total effective throughput of all $\psi$ configuration is $y_\psi \cdot h_{\psi}$ (noted that $h_{\psi}$ denotes the per-replica throughput of configuration $\psi$ defined before). Consequently, the time required for configuration $\psi$ to complete its assigned workloads is $\Theta_\psi = \frac{ x_\psi \cdot \text{len}_{\text{rollout}}}{y_{\psi} \cdot h_{\psi}}$. Since all rollout inference replicas operate in parallel, the system completes when the slowest configuration finishes, which imposes the constraint $\Theta_\psi \leq \Theta$ for all $\psi \in \Psi$. Our objective is to minimize the makespan $\Theta$.

\textbf{MILP formulation.} The problem for optimizing rollout generation throughput can be summarized as the following Mixed-Integer Linear Program (MILP):
\vspace{-0.5em}

\begin{equation}
\begin{aligned}
\arg&\min \quad \Theta \label{eq:obj}\\
\text{s.t.} &\enspace \sum_{\psi \in \Psi} x_{\psi} = B, \frac{x_{\psi} \cdot \text{len}_{\text{rollout}} }{y_\psi \cdot h_{\psi,w}} \leq \Theta, \enspace \forall \psi \in \Psi, \\
&\enspace \sum_{\psi \in \Psi} i^\psi_t \cdot y_\psi \leq i_t, \enspace \forall t\in [1, T].
\end{aligned}
\end{equation}%

After solving this MILP, we obtain a rollout generation execution plan $\tau$, which encompasses replica counts for each configuration ($y_\psi$) and workload assignment ($\mathcal{A}_I$) through the above formulation, subject to memory and GPU availability constraints. To accelerate the algorithm, we restrict rollout replicas to use bandwidth-sensitive tensor model parallelism only within single machines to reduce the search space. Additionally, since reward computation varies minimally across rollout generation tasks, we treat the reward computation cost as a constant obtained through profiling, which would not influence MILP results.

\textbf{Optimize across different $\delta(\eta)$ values.} 
Across algorithm iterations, we iteratively optimize parallel strategy and workload assignment for model training and rollout generation under varying step window $\delta(\eta)$. The choice of $\delta(\eta)$ involves a fundamental tradeoff: optimizing over the entire training horizon ($\delta=\infty$) is computationally infeasible and requires an unknown total training iteration, while optimizing over too few steps ($\delta \approx 1$) fails to accurately capture asynchronous RL training and data staleness characteristics. To balance these concerns, we adopt an adaptive strategy: We initialize $\delta(\eta)$ to a small value empirically determined from the data staleness parameter $\eta$, then iteratively increase $\delta(\eta)$ until the scheduling results exhibit stable RL training execution latency. This approach leverages the key insight that when $\delta(\eta)$ accurately captures the asynchronous training dynamics and data staleness characteristics, scheduling decisions remain stable under larger values of $\delta(\eta)$, indicating diminishing returns from further increases of $\delta(\eta)$.

\subsection{Repartition Phase Algorithm}
\label{sec:second-phase}

The previous search-phase presented an approach for obtaining the
optimal execution plans for a given resource allocation. On the other hand, in this section, we introduce a cost-guided graph partition algorithm noted by $\textsc{Graph\_Partition}(C_T, C_I, \mathcal{D})$ to find a resource allocation that minimizes the overall execution cost of the system.

\textbf{Cost-guided graph partition.} To determine the optimal resource allocation of the GPU set $\mathbf{D}$ for model training $\mathcal{D}_T$ and the set of GPUs for rollout generation $\mathcal{D}_I$, we design a graph partition approach guided by the costs generated in the first phase. We extend the GPU set $\mathcal{D}$ to a graph $G=(\mathcal{D}, \mathcal{E})$, where each GPU $d \in \mathcal{D}$ is a vertex noted with a computation power $c_d$, HBM memory bandwidth $m_d$, and HBM limit $M_d$; $\forall d_1, d_2 \in \mathcal{D}$, communication bandwidth $\beta_{d_1, d_2}$ represents an edge between GPUs in set $\mathcal{E}$. We bisect graph $G$ into partition $\mathcal{D} = \{\mathcal{D}_T, \mathcal{D}_I \}$, where $\mathcal{D}_T \cup \mathcal{D}_I=\mathcal{D}$ and $\mathcal{D}_T \cap \mathcal{D}_I=\emptyset$. When partitioning the graph, we adopt the following heuristics:

\begin{itemize}

    \item \textit{High network bandwidth should be allocated for model training.} Since model training involves intensive collective communication operations (e.g., \texttt{all-reduce} in TP and DP) that demand high communication volume~\cite{zhong2025streamrl}, we prioritize allocating GPUs with high aggregate interconnect bandwidth to model training. Formally, we define the metric $\frac{\sum_{d_1, d_2 \in \mathcal{D}_T} \beta_{d_1, d_2}}{\sum_{d_1', d_2' \in \mathcal{D}} \beta_{d_1', d_2'}}$, i.e., the aggregated communication bandwidth of $\mathcal{D}_T$ over the aggregated communication bandwidth of $\mathcal{D}$, to guide the graph partition by this heuristic.
    
    \vspace{-0.25em}
    \item \textit{High HBM bandwidth GPUs should be allocated for rollout generation.} Since rollout generation is mainly HBM I/O-bound during generative inference computation, we prioritize allocating GPUs with high HBM bandwidth to rollout generation. Formally, denoting  we define the metric $\frac{\sum_{d\in \mathcal{D}_I } m_d}{\sum_{d'\in \mathcal{D} } m_{d'}}$, i.e. the aggregated HBM bandwidth of $\mathcal{D}_I$ over the aggregated HBM bandwidth of $\mathcal{D}$, to guide the graph partition.
    \vspace{-0.25em}

    \item \textit{Careful balance of the computation power should maintained between model training and rollout inference.} Both rollout generation and model training demands the computation capacity, to maintain an adaptive balance, we let the fraction of computation capacity occupied by model training, i.e., $\frac{\sum_{d\in \mathcal{D}_T} c_d}{\sum_{d'\in \mathcal{D}} c_{d'}}$, to fall into a tunable range $[ \gamma_L, \gamma_H]$, where $\gamma_L$ and $\gamma_H$ will be tuned automatically along with the iterative resource allocation as we will enumerate next. 
\end{itemize}%

By integrating the above criteria into our graph partition formulation, we define the final constrained optimization as:%

\begin{equation}
\begin{aligned}
\mathcal{D}^{*}_T, \mathcal{D}^{*}_I = \arg \max_{\mathcal{D}_T,  \mathcal{D}_I } \quad & \left(  \frac{\sum_{d_1, d_2  \in \mathcal{D}_T} \beta_{d_1, d_2}}{\sum_{d'_1, d'_2  \in \mathcal{D}} \beta_{d'_1, d'_2}} + \frac{ \sum_{d\in \mathcal{D}_I} m_d }{\sum_{d'\in \mathcal{D}} m_{d'}}\right) \\
s.t.  \quad & \gamma_L\leq \frac{\sum_{d\in \mathcal{D}_T} c_d}{\sum_{d'\in \mathcal{D}} c_{d'}} \leq \gamma_H
\end{aligned}
\end{equation}%

\textbf{Iterative refinement.} Across graph partition optimization iterations, we dynamically adjust the fraction of total computational power allocated to model training (i.e., $\gamma_L$ and $\gamma_H$) based on the execution costs generated in the search phase algorithm (i.e., $C_T$ and $C_I$). To balance model training and rollout generation costs, we fine-tune $\gamma_L$ and $\gamma_H$, which control the computational power ratio between them. We initialize $\gamma_L = \gamma_H = 1$, assigning all FLOPS to model training. In subsequent iterations, the optimal resource allocation is determined by tuning $\gamma_L$ and $\gamma_H$ using a binary search procedure. Specifically, we define the search range as $[q, r]$, where $q=0$ and $r=1$ initially. At each iteration, we first update the FLOPS allocated to model training by updating $\gamma_L = \gamma_H = (q + r)/2$, and then evaluate the execution costs $C_T$ and $C_I$: If $C_T < C_I$, we set the upper bound $r = (q + r)/2$;
otherwise, if $C_T \geq C_I$, we update the lower bound $q = (q + r)/2$.
This process continues until $C_T \approx C_I$, at which point the final resource allocation is selected to minimize the overall execution cost.

\label{sec:eval}
This section presents a comprehensive evaluation of \sys across various configurations. We aim to investigate the following research questions:
\begin{itemize}

\item \textit{RQ1: What is the end-to-end performance of \sys under heterogeneous clusters and \sysh under homogeneous clusters with the same total budget?}
\vspace{-0.25em}
\item \textit{RQ2: What is the breakdown performance of \sys and \sysh, and how does each RL training component contribute to the RL training step execution latency?}
\vspace{-0.25em}
\item \textit{RQ3: What is the ablation performance and cost efficiency of \sys and \sysh?}
\vspace{-0.25em}
\item \textit{RQ4: How effective is the scheduling algorithm of \sys?}
\end{itemize}

\subsection{Experimental Setup}
\label{subsec:setup}
\textbf{Experimental environments.} Our heterogeneous RL training experiments are conducted using two distinct GPU architectures: NVIDIA H20 GPUs\cite{h20}, which provide 148 TFLOPS of tensor core computational power, 4 TB/s memory bandwidth, and 450 GB/s single-direction NVLink intra-machine bandwidth; and NVIDIA H800 GPUs\cite{h800specs}, which deliver 756 TFLOPS of tensor core computational power, 2 TB/s memory bandwidth, and 200 GB/s single-direction NVLink intra-machine bandwidth. The inter-machine bandwidth between H20 or H800 machines is 5 GB/s, while the communication bandwidth between H20 and H800 machines is 1.5 GB/s. The homogeneous RL training experiments are conducted using homogeneous NVIDIA H800 GPUs or H20 GPUs.

\textbf{Baselines.} Since our primary focus is accommodating RL training on heterogeneous GPUs, we primarily compare the asynchronous RL training end-to-end performance of \sys in heterogeneous environments against \sysh in homogeneous environments under equivalent total budgets. For RL training algorithms, we evaluate \sys and \sysh on the GRPO algorithm, which is among the most impactful RL algorithms in current practice. We employ identical training hyperparameters to those used in \sysh~\cite{fu2025areal}.

\textbf{Evaluation metrics.} Following established practices in prior research~\citep{fu2025areal}, we employ the RL training throughput metric to assess system efficiency. RL training throughput is calculated by dividing the average number of tokens processed by the average RL training execution time, which is defined as the sum of the weight synchronization latency and the maximum duration between rollout generation (encompassing both rollout and reward computation) and model training (encompassing policy and, optionally, value model gradient computation). Given that RL training with \sys and \sysh operates asynchronously, we compute the RL training throughput as the average value across 30 RL training steps.

\begin{figure}
    \centering
    \includegraphics[width=0.6\linewidth]{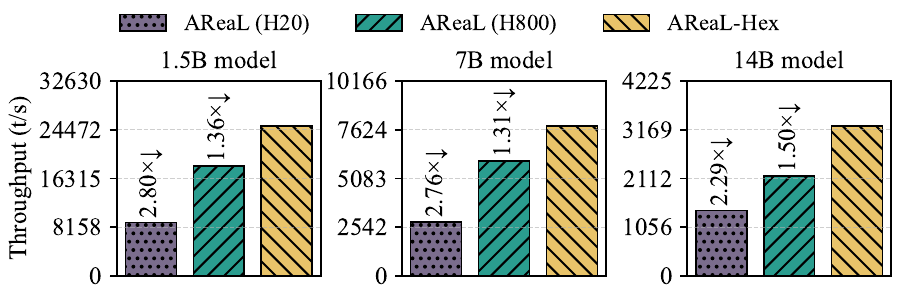}
    
    \caption{We present the end-to-end experimental results of \sys operating on H800 and H20 mixed heterogeneous GPUs compared to \sysh operating on homogeneous H800 or H20 GPUs. All experimental settings maintain the same total budgets using the same prices as in \S\ref{sec:case}.}
    
    \label{fig:e2e}
\end{figure}

\subsection{End-to-end System Performance Comparison (Q1)}
\label{sec:e2e}

\textbf{End-to-end experimental results:} On the GRPO RL training workloads, we benchmark the end-to-end training performance of \sys with heterogeneous clusters against \sysh with homogeneous clusters (either homogeneous H800 GPUs or H20 GPUs) under the same total budget. We compare the asynchronous RL training throughput of \sys and \sysh. As shown in \autoref{fig:e2e}, experimental results across different model sizes (ranging from 1.5B to 14B parameters) of the DeepSeek Distilled Qwen-2.5 models~\cite{deepseek-r1-distill-qwen-1.5b,deepseek-r1-distill-qwen-7b,deepseek-r1-distill-qwen-14b} demonstrate that \sys consistently reduces RL training latency across all evaluated cases when both \sys and \sysh apply the optimal parallel configurations. Compared to the baseline where \sysh runs on H800 homogeneous GPUs, \sys achieves up to 1.50$\times$ speedup (on the 14B model), at least 1.31$\times$ speedup (on the 7B model), and an average 1.39$\times$ speedup; compared to the baseline where \sysh runs on H20 homogeneous GPUs, \sys achieves up to 2.76$\times$ speedup (on the 7B model), at least 2.29$\times$ speedup (on the 14B model), and an average 2.62$\times$ speedup.

\textbf{Discussion:} In essence, heterogeneous RL training aligns diverse hardware capabilities with the heterogeneous computational demands of RL training. By strategically allocating H20 GPUs for memory-bound rollout generation and H800 GPUs for compute-bound model training, we effectively leverage the advantages of both GPU architectures. H800 GPUs excel at computation-intensive tasks, while H20 GPUs demonstrate superior performance in rollout generation due to their distinct hardware specifications on computational power and HBM memory bandwidth. Our experimental results demonstrate that \sys deployed on heterogeneous GPUs consistently outperforms \sysh running on either homogeneous H800 or H20 GPU clusters. Specifically, the H800 homogeneous baseline exhibits suboptimal performance in rollout generation, whereas the H20 homogeneous baseline underperforms in model training phases. These findings confirm that strategic heterogeneous deployment, which respects the distinct computational characteristics of different hardware architectures, yields superior overall performance.

\begin{figure}
    \centering
    \includegraphics[width=0.6\linewidth]{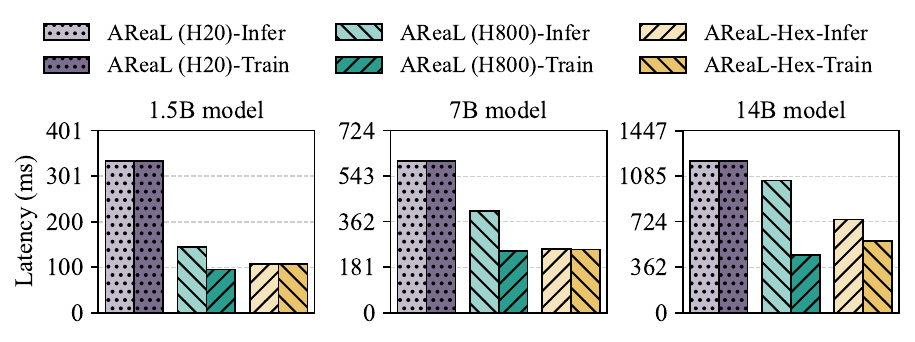}
    
    \caption{We present a breakdown of experiments comparing \sys running on a 56-GPU heterogeneous cluster against \sysh running on a 24-GPU H800 homogeneous cluster. The terms \textsc{Infer} and \textsc{Train} denote the execution latencies for rollout generation and model training.}
    \label{fig:bk}
\end{figure}

\subsection{Breakdown System Performance Comparison (Q2)}
\label{sec:bk down}

\textbf{Performance breakdown results.} We present a comprehensive breakdown performance of the asynchronous GRPO RL training using \sys (running over heterogeneous GPUs) and \sysh (running over homogeneous H800 or H20 GPUs) across various model sizes in \autoref{fig:bk}. Compared to \sysh running over homogeneous H800 GPUs, \sys achieves a maximum of 1.61$\times$, a minimum of 1.35$\times$, and on average 1.46$\times$ lower rollout generation execution latency; compared to \sysh running over homogeneous H20 GPUs, \sys achieves a maximum of 3.13$\times$, a minimum of 1.85$\times$, and on average 2.46$\times$ lower model training execution latency. The comparison of weight synchronization execution cost, which comprises a small fraction for both \sys and \sysh, is presented in \autoref{tab:update_time}. For \sysh operating on H800 GPUs, weight synchronization costs represent, on average, 1.9\% of the GRPO RL training iteration duration. For \sys operating in heterogeneous GPU environments with 1.5 GB/s communication bandwidth, weight synchronization costs account for 13.54\% of the total RL training time.

\begin{table}[ht]
\centering

\caption{Weight update time comparison across homogeneous and heterogeneous GPU configurations.}

\label{tab:update_time}
\resizebox{0.6\columnwidth}{!}{%
\begin{tabular}{l | c | c | c}
\hline
\textbf{Model} & \textbf{\sysh (H800)} & \textbf{\sysh (H20)} & \textbf{\sys} \\
\hline
\hline
\textbf{1.5B model} & 4.75 s & 2.74 s & 10.06 s \\
\hline
\textbf{7B model}   & 14.79 s & 7.46 s & 58.34 s \\
\hline
\textbf{14B model}  & 26.00 s & 13.05 s & 112.93 s \\
\hline
\end{tabular}%
}
\end{table}

\textbf{Discussion:} Experimental results demonstrate that rollout generation in \sysh running on homogeneous H800 GPUs frequently emerges as the primary system bottleneck, as the relatively short model training duration fails to provide sufficient temporal overlap with rollout generation. In contrast, \sys running on heterogeneous GPUs exhibits more balanced rollout generation and model training speeds, achieving optimal system performance under asynchronous operation. Fundamentally, the integration of H20 GPUs enhances rollout generation capacity, thereby delivering substantial system performance improvements. Moreover, the communication overhead between heterogeneous devices remains acceptable under 1.5GB/s inter-machine bandwidth constraints. Given that 1.5GB/s communication bandwidth between heterogeneous GPUs is typically readily available in practice, heterogeneous RL training deployment represents a practical solution for real-world applications without requiring extensive engineering efforts for network communication across heterogeneous devices.

\begin{table}[h]
\centering
\caption{Throughput comparison between \sysh and \sysh with uniform resource allocation, i.e., \sysh (u).}

\label{tab:resource}
\resizebox{0.6\columnwidth}{!}{%
\begin{tabular}{l | c | c}
\hline
\textbf{Model} & \textbf{\sysh\ (u)} & \textbf{\sysh} \\
\hline
\hline
\textbf{1.5B model} & $1.17\times10^4$ t/s & $1.84\times10^4$ t/s ($1.57\times$$\uparrow$) \\
\hline
\textbf{7B model}   & $3.57\times10^3$ t/s & $5.98\times10^3$ t/s ($1.68\times$$\uparrow$) \\
\hline
\textbf{14B model}   & $1.29\times10^3$ t/s & $2.16\times10^3$ t/s ($1.67\times$$\uparrow$) \\
\hline
\end{tabular}%
}
\end{table}

\begin{figure}
    \centering
    \includegraphics[width=0.6\linewidth]{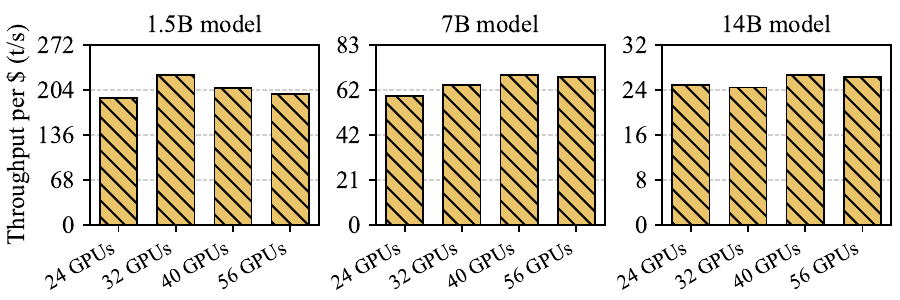}
    
    \caption{Case study of \sys's performance cost-efficiency across different cluster sizes ranging from 24 to 56 GPUs. For the H20 and H800 GPU per-hour costs, we follow the prior practice~\cite{zhu2025megascale}.}
    
    \label{fig:cost-efficiency}
\end{figure}

\begin{table}[h]
\centering
\caption{Cost-efficiency comparison between \sys running on a 32-GPU heterogeneous cluster and \sysh running on 24 H800 GPUs.}
\label{tab:cost-eff-compare}

\resizebox{0.6\columnwidth}{!}{%
\begin{tabular}{l | l | c | c}
\hline
\textbf{Model} & \textbf{System} & \textbf{Throughput} & \textbf{Cost} \\
\hline
\hline
\multirow{2}{*}{\textbf{1.5B model}} & \sysh (H800) & $1.84\times10^4$ t/s & \$126.72/h \\
\cline{2-4}
 & \sys & $1.79\times10^4$ t/s & \$86.64/h \\
\hline
\multirow{2}{*}{\textbf{7B model}} & \sysh (H800) & $5.98\times10^3$ t/s & \$126.72/h \\
\cline{2-4}
 & \sys & $6.00\times10^3$ t/s & \$86.64/h \\
\hline
\multirow{2}{*}{\textbf{14B model}} & \sysh (H800) & $2.16\times10^3$ t/s & \$126.72/h \\
\cline{2-4}
 & \sys & $2.30\times10^3$ t/s & \$86.64/h \\
\hline
\end{tabular}%
}
\end{table}

\subsection{Ablation Study \& Case Studies (Q3)}
\label{sec:ablation}

\textbf{Ablation on resource allocation.} In \autoref{tab:resource}, we present the performance variations of \sysh under different resource allocation strategies. The experimental results demonstrate that our resource allocation optimization approach (detailed in \S\ref{sec:second-phase}) delivers substantial and consistent improvements in system performance for 1.5B, 7B, and 14B models. Specifically, our resource allocation optimization achieves a maximum speedup of 1.68$\times$, and an average speedup of 1.63$\times$ compared to configurations where this scheduling optimization is disabled. These results highlight the necessity of careful scheduling of the resource allocation.

\textbf{Case study on the cost efficiency of \sys.} In \autoref{fig:cost-efficiency}, we present the per-dollar end-to-end system performance of \sys under varied cluster sizes. Experimental results demonstrate that \sys maintains stable per-dollar system throughput across heterogeneous GPU clusters ranging from 24 to 56 GPUs. Concretely, the per-dollar throughput for the 1.5B, 7B, and 14B models is around 200, 62, and 24 tokens/s, across cluster configurations ranging from 24 to 56 heterogeneous GPUs. These results validate that heterogeneous RL training can deliver stable performance, regardless of the cluster size.

\textbf{Case study on the cost efficiency comparison.} In \autoref{tab:cost-eff-compare}, we present the experimental results where \sys and \sysh achieves the same throughput. In this case, \sys running over heterogeneous GPUs demonstrates a lower per-hour training cost than \sysh running over homogeneous GPUs. For 1.5B, 7B, and 14B models, the heterogeneous setting demonstrates 1.42$\times$, 1.31$\times$, and 1.50$\times$ lower per-hour training cost than the H800 homogeneous baseline. For the H20 baseline, our heterogeneous settings achieving the same throughput also demonstrate a 50\% RL training cost reduction. These results confirm that heterogeneous RL training is an economical choice.

\subsection{Evaluation on Scheduling Algorithm (Q4)}
\label{sec:eval-algo}

\begin{table}[h]
\centering

\caption{Scheduling (algorithm convergence) time across different cluster sizes. ``Ours (w/o Search)" and ``Ours (w/o Repartition)" mean replacing the search phase (\S\ref{sec:first-phase}) and the repartition phase (\S\ref{sec:second-phase}) of our scheduling algorithm with exhaustive search, respectively. The algorithm is considered converged when the solution remains stable for more than 20 consecutive iterations.}

\resizebox{0.7\columnwidth}{!}{
\begin{tabular}{l|c|c|c}
\hline
\textbf{Cluster Size} & \textbf{Ours} & \textbf{Ours (w/o Search)} & \textbf{Ours (w/o Repartition)} \\
\hline
\hline
\textbf{24 GPUs} & 14.9s & 6.1min (24.6$\times$$\uparrow$) & 5.0min (20.1$\times$$\uparrow$) \\
\hline
\textbf{32 GPUs} & 23.1s & 11.3min (29.4$\times$$\uparrow$) & 7.7min (20.0$\times$$\uparrow$) \\
\hline
\textbf{40 GPUs} & 46.7s & 34.4min (44.2$\times$$\uparrow$) & 16.4min (21.1$\times$$\uparrow$) \\
\hline
\textbf{56 GPUs} & 2.0min & $\geq$ 40min ($\geq$ 20.0$\times$$\uparrow$) & $\geq$ 40min ($\geq$ 20.0$\times$$\uparrow$) \\
\hline
\end{tabular}
}
\label{tab:algo-speed}
\end{table}

\textbf{Algorithm execution speed.} In \autoref{tab:algo-speed}, we present a comparative analysis of the execution speed of our scheduling algorithm against two baselines where each of our algorithm phases is replaced with an exhaustive search approach. Experimental results demonstrate that our algorithm efficiently determines optimal resource allocation and parallelization strategies for rollout generation and model training, achieving a substantial speedup of 20.0$\times$ to 44.2$\times$ in search time compared to the exhaustive search baseline algorithms.

\section{Conclusion}

In this paper, we introduced \sys, a novel system that enhances asynchronous RL training on heterogeneous GPUs. We carefully designed a two-phase scheduling algorithm, which is capable of effectively and efficiently determining the resource allocation and parallel strategy for rollout generation and model training. Maintaining the same total, experimental results demonstrate that \sys running over heterogeneous GPUs achieves superior performance than \sysh running over homogeneous GPUs on models with 1.5B, 7B, and 14B model parameters. By allocating suitable resources for rollout generation and model training, \sys has unlocked the potential of heterogeneous resources and provides an economic alternative compared to the homogeneous clusters.

\bibliographystyle{unsrt}
\bibliography{main}

\end{document}